# Photon statistics of incoherent cathodoluminescence with continuous and pulsed electron beams

Magdalena Solà-Garcia,[1,*] Kelly W. Mauser,[1] Matthias Liebtrau,[1] Toon Coenen,[1,2] Sophie Meuret,[3] Albert Polman[1]

[1] Center for Nanophotonics, AMOLF, Science Park 104, 1098 XG Amsterdam, The Netherlands
[2] Delmic BV, Kanaalweg 4, 2628 EB, Delft, The Netherlands
[3] CEMES-CNRS, 29 Rue Jeanne Marvig, 31055 Toulouse, France

[*] Corresponding author: m.sola@amolf.nl

**ABSTRACT:** Photon bunching in incoherent cathodoluminescence (CL) spectroscopy originates from the fact that a single high-energy electron can generate multiple photons when interacting with a material, thus revealing key properties of electron-matter excitation. Contrary to previous works based on Monte-Carlo modelling, here we present a fully analytical model describing the amplitude and shape of the second order autocorrelation function ($g^{(2)}(\tau)$) for continuous and pulsed electron beams. Moreover, we extend the analysis of photon bunching to ultrashort electron pulses, in which up to 500 electrons per pulse excite the sample within a few picoseconds. We obtain a simple equation relating the bunching strength ($g^{(2)}(0)$) to the electron excitation efficiency ($\gamma$), electron beam current, emitter decay lifetime and pulse duration, in the case of pulsed electron beams. The analytical model shows good agreement with experimental data obtained on InGaN/GaN quantum wells using continuous, ns-pulsed (using beam blanker) and ultrashort ps-pulsed (using photoemission) electron beams. We extract excitation efficiencies of 0.13 and 0.05 for 10 and 8 keV electron beams, respectively, and we observe that non-linear effects play no compelling role even after excitation with ultrashort and dense electron cascades in the quantum wells.

Photon statistics in incoherent cathodoluminescence (CL) reveals fundamental properties of the interaction of high-energy electrons (~1-300 keV) with matter[1]. In particular, the second-order autocorrelation function ($g^{(2)}(\tau)$) exhibits strong bunching ($g^{(2)}(0) \gg 1$) when exciting a material, such as a semiconductor or insulator, with an electron beam[1–3], contrary to conventional photoluminescence measurements with laser excitation (typically $g^{(2)}(0) = 1$)[4]. This is because each electron initially excites bulk plasmons in the material, which end up generating thermalized carriers that diffuse and recombine. This recombination can lead to either the emission of a photon with energy corresponding to the bandgap of the material (bimolecular recombination), or to the excitation of another emitter embedded in the material, such as a defect or quantum well, which can then decay radiatively. Either cases can result in the emission of multiple photons per incoming electron[5]. Recently, $g^{(2)}(\tau)$ measurements have been used to quantify the excitation efficiency of electrons in InGaN/GaN quantum wells (QWs) with different geometries[6,7]. Moreover, the $g^{(2)}(\tau)$ technique allows to extract the emitter decay dynamics without the need of a pulsed electron beam[1].

These new insights into the use of $g^{(2)}(\tau)$ measurements in CL are key for a complete analysis of electron microscopy experiments.

All CL bunching experiments performed so far have been described using Monte Carlo-based numerical models, showing good agreement with the measured $g^{(2)}(\tau)$ curves and the dependence of $g^{(2)}(0)$ on the electron current. However, Monte Carlo models are time consuming and fail to provide a full understanding of the bunching process and the key parameters that determine its amplitude. Moreover, fitting the experimental data with a Monte Carlo model is complex and requires additional computation and interpolation procedures, thus making the $g^{(2)}(\tau)$ analysis less accessible.

In addition to this, CL autocorrelation measurements so far have been limited to the cases of continuous and ns-pulsed electron beams. Recently, ultrafast electron microscopy using fs-ps electron pulses as excitation sources, has emerged as a powerful technique to access the dynamics of electron excitation of materials with high temporal resolution, combined with the nanoscale electron-beam spatial resolution[8–10]. Ultrafast electron microscopy has already been used to study electron-generated carrier dynamics[11,12] and phase transitions[13,14], among others. Additionally, the development of techniques such as photon-induced near-field microscopy (PINEM) has exploited the quantum nature of the electron wave packet[15,16], thus leading the way to the study of quantum-mechanical aspects of electron-light-matter interactions inside an electron microscope. Autocorrelation measurements, such as $g^{(2)}(\tau)$, using ultrashort electron pulses can offer new insights into the dynamics of excitation of a material with dense electron pulses[17].

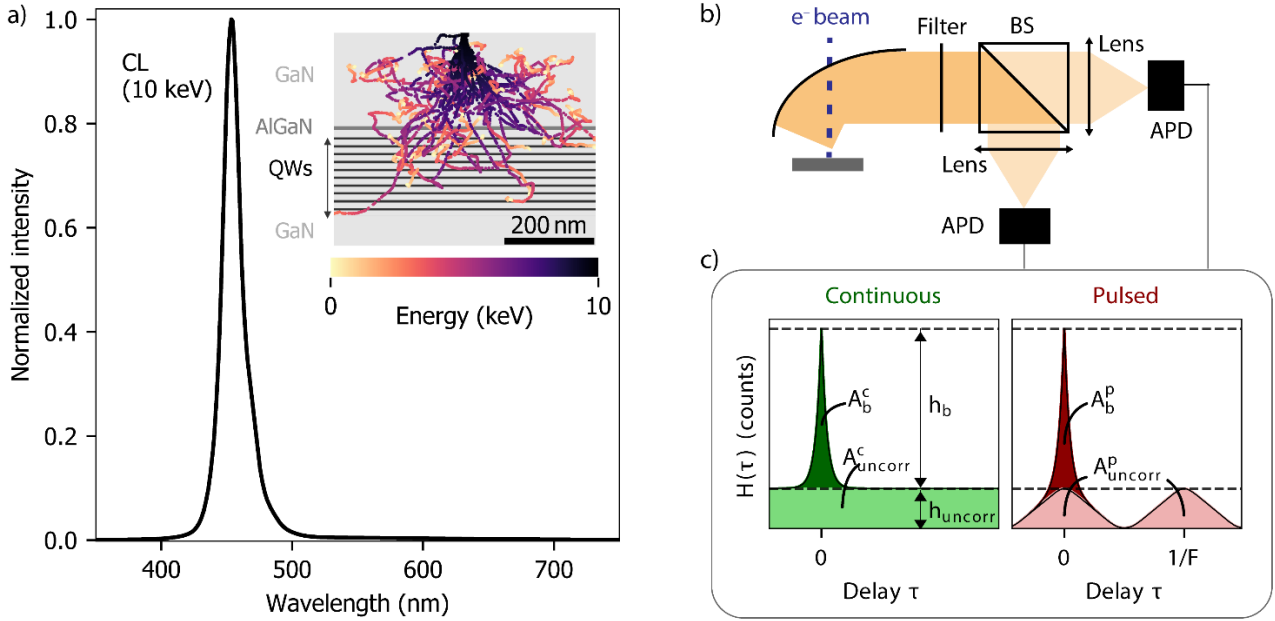

**Figure 1:** (a) Cathodoluminescence (CL) spectrum of InGaN/GaN quantum wells (QWs) obtained with a continuous 10 keV beam (242 pA). Inset: schematic of the InGaN/GaN heterostructure overlaid with the results from Casino simulations. Each curve indicates the trajectory of one electron, and the color bar indicates the energy of the electron at each position. (b) Schematic of the CL collection and analysis using a Hanbury-Brown and Twiss (HBT) experiment. (c) Schematic of the expected histograms obtained using the HBT experiments using continuous (left) and pulsed (right) electron beams.

In this paper, we resolve the above-mentioned limitations of current $g^{(2)}(\tau)$ analyses which use Monte Carlo simulations. We develop a fully analytical model to describe the value of $g^{(2)}(0)$ as a function of four experimental parameters, for different electron beam configurations. Our analytical model describes how the electron current (or number of electrons per pulse), emitter lifetime, excitation efficiency and pulse duration, in the case of pulsed electron beams, determine the value of $g^{(2)}(0)$. Using our analysis, we directly extract the electron excitation efficiency $\gamma$, defined as the fraction of electrons that create at least one interaction with the emitter[6], from one simple equation. We also show that our model reproduces the Monte Carlo simulations developed in previous work.

In order to further test the validity of the model, we perform $g^{(2)}(\tau)$ experiments on InGaN/GaN quantum wells with both continuous and pulsed electron beams. In particular, we study two types of pulsed electron beams: with relatively long (up to 200 ns) and ultrashort (a few ps) pulse durations. In the case of ultrashort pulses, we vary the number of electrons per pulse from (on average) less than 1 up to ~500, thus allowing us to access regimes in which several electrons interact with the sample within the bulk plasmon decay and carrier thermalization timescales. Here, our analytical model shows that $g^{(2)}(0)$ depends only on the number of electrons per pulse and the excitation efficiency. From the model it can also be derived that in the case of a pulsed electron beam the excitation efficiency can be obtained alternatively through a simple analysis, without the need of any fitting procedure. Our analysis of the $g^{(2)}(\tau)$ function shows that even for dense cascades generated by 500 electrons per pulse (i.e., within a few picoseconds) nonlinear effects do not have a compelling contribution in the excitation and carrier recombination of InGaN/GaN QWs.

*Experimental section*

Cathodoluminescence experiments are performed in a scanning electron microscope (SEM) equipped with a parabolic mirror that collects the emitted light. The statistics of CL emission is analyzed using a Hanbury-Brown and Twiss (HBT) geometry[18], composed of a 50:50 beam splitter (BS) and two avalanche photodiodes (APDs) as single-photon counting detectors (Fig. 1b). Experiments with varying pulse widths (6 − 200 ns) are performed using an electrostatic electron beam blanker. Ultrashort pulses, with pulse widths of few picoseconds, are ob-tained by focusing a fs-laser ($\lambda = 258$ nm) onto the electron cathode, inducing photoemission of electron pulses[19,20].

We study a bulk semiconductor heterostructure of InGaN/GaN quantum wells, grown by molecular beam epitaxy[6]. A schematic of the structure is shown in the inset of Fig. 1a. The sample consists of 10 2-nm-thick InGaN layers, separated with 15-nm GaN layers. A 2-nm AlGaN barrier layer is grown on top of the quantum well stack, and the whole structure is buried below a 250-nm-thick p-type GaN layer. The substrate is composed of n-type GaN. The inset also shows the results of a Casino simulation using a 10 keV electron beam[21]. Each dot in the plot represents a collision of the primary electron with the sample, and thus the creation of one or more bulk plasmons. The color of the dot indicates the energy of the primary electron beam at that point. The results show that only a small fraction of the electrons will directly reach the QWs, as previously shown using $g^{(2)}(\tau)$ measurements[6]. Moreover, the AlGaN layer acts as a carrier blocking layer[22], hence only carriers generated below this layer can excite the QWs.

Fig. 1a shows a typical CL spectrum obtained when exciting the sample with 10 keV electrons. The emission originates mostly from the QWs, corresponding to the InGaN band edge emission peak around 450 nm. Defect luminescence from the yellow band[23,24], in the 520-650 spectral range, is barely visible in the spectrum, given that the intensity in this range is 30 times lower than the QW emission. This is in accordance with previous CL measurements on this sample[6], and is attributed to the fact that 10 keV electrons do not reach the GaN substrate, thus limiting the excitation of carriers in the bulk GaN. In the HBT experiments we use a bandpass filter (450±40 nm) to ensure that only the CL emission from the QWs is recorded.

*Continuous electron beam*

A typical $g^{(2)}(\tau)$ experiment consists on the acquisition of a histogram ($H(\tau)$) of the number of coincidence events (i.e., a correlation) as a function of the delay between two recorded photons ($\tau$). As schematic of the recorded histogram obtained after excitation with a continuous electron beam is shown in Fig. 1c (left). In this case, the $g^{(2)}(\tau)$ curve is obtained by normalizing the histogram with respect to the value at very long delay ($\tau \to \infty$), $h_{\text{uncorr}}$, which represents the amplitude for uncorrelated events. Hence,

$$g^{(2)}(0) = \frac{H(0)}{H(\tau \to \infty)} = 1 + \frac{h_b^c}{h_{\text{uncorr}}^c}, \quad (1)$$

where $h_b^c$ is the height of the bunching peak, as depicted in Fig. 1c. $g^{(2)}(0)$ can be interpreted as the likelihood of having two photons with delay $\tau = 0$ compared to any other delay. In the case of Poissonian statistics, such as for coherent light, $g^{(2)}(\tau) = 1$ for any delay[4], while $g^{(2)}(0) < 1$ indicates sub-Poissonian statistics (anti-bunching), as in the case of a single-photon emitter[25,26], while $g^{(2)}(0) > 1$ represents super-Poissonian statistics (bunching). Some examples of processes in which bunching occurs are blackbody radiation[27,28] and superradiance[29,30], as well as the CL emission presented here. From a statistical point of view, $h_b$ is related to the total number of correlations (i.e., detection of two photons) leading to bunching, i.e., coming from the same electron, while $h_{\text{uncorr}}^c$ represents the uncorrelated events, i.e., correlations between photons that are generated by different electrons.

The temporal decay of the bunching peak is determined by the radiative decay of the emitter, and enables determination of the emitter lifetime, as will be explained below. The area of the bunching peak, related to the height as $A_b^c = \alpha_b h_b^c$, is proportional to the average number of possible combinations between pairs of photons that lead to bunching, i.e., that come from the same electron. Here we have defined $\alpha_b$ as the shape factor of the bunching peak. Similarly, the area $A_{\text{uncorr}}^c$ is related to the mean number of combinations of photon correlations from different electrons, i.e., uncorrelated events, during the acquisition time of the experiment $T = Bt_B$. Here $B$ is the total number of bins in the experiment and $t_B$ is the time of each bin. From this it follows that $A_{\text{uncorr}}^c = h_{\text{uncorr}}^c (2B + 1) t_B / 2$. The factor $2B + 1$ comes from the fact that the $g^{(2)}(\tau)$ histogram is theoretically built over positive and negative times, in a symmetric fashion. The additional factor 2 in the denominator accounts for the fact that the number of possible events decreases for increasing delay following a triangular function (see S2f). Taking these definitions into account, Eq. (1) becomes

$$g_{cont}^{(2)}(0) = 1 + \frac{A_b^c}{A_{\text{uncorr}}^c} \frac{(2B+1)t_B}{2\alpha_b}. \quad (2)$$

The model is constructed following the subsequent steps that start with an electron entering the material, until a photon is emitted, similar to the previous Monte Carlo model[1,6]. The steps are:

1. Excitation of $b_i$ bulk plasmons close to the QWs, described by a Poisson distribution with expected value $b$. It should be noted that the number of plasmons generated per electron will probably be larger than $b_i$, but here we only consider those effectively exciting the QWs.

2. Decay of each bulk plasmon into $m_i$ thermalized electron-hole pairs, again described by a Poisson distribution with expected value $m$.

3. Diffusion of carriers, which can either:
   a. excite a QW, which emits a photon, with probability $\eta$.
   b. not excite a QW or excite a QW which then decays non-radiatively.

This step is assumed to follow a binomial distribution, with $m_i$ representing the number of events and $\eta$ probability that an event results in the emission of a photon.

A key aspect of our $g^{(2)}(0)$ model is that it accounts for the combined stochastic nature of all the involved processes. The model is therefore based on the calculation of the average possible combinations of correlations that lead to bunching ($A_b^c$) and to uncorrelated events ($A_{\text{uncorr}}^c$). A full derivation of the model is provided in the Supporting Information (S2). In brief, from step 2 and 3 we obtain that the average number of possible combinations of correlations created after the excitation of $b_i$ bulk plasmons is given by $b_i^2 m^2 \eta^2$. We then need to find the average number of combinations of correlations between photons from the same electron (i.e. ignoring combinations of correlations created by photons from different electrons), taking into account that $b_i$ follows a Poisson. Given $n$ electrons arriving at the sample during the time of the experiment $T$, the average number of combinations of correlations leading to bunching becomes

$$A_b^c = n\, b(b+1) m^2 \eta^2. \tag{3}$$

Similarly, it can be shown that the expected value of the number of combinations of correlations leading to uncorrelated events, i.e., pairs of photons coming from different electrons, is (see S2b)

$$A_{\text{uncorr}}^c = n(n-1) b^2 m^2 \eta^2. \tag{4}$$

We now insert Eqs. (3) and (4) into Eq. (2), and rewrite $n$ as a function of the electron current, $I = nq/(Bt_b)$, where $q$ is the electron charge. We also consider the limit $B \gg 1$, which is reasonable given that the acquisition time is typically minutes or more while the time resolution is usually less than 1 ns. We then obtain the following expression for the amplitude of $g^{(2)}(\tau)$ at 0 delay:

$$g_{\text{cont}}^{(2)}(0) = 1 + \frac{q}{I\alpha_b}\frac{b+1}{b}. \tag{5}$$

Several aspects are noticeable from Eq. (5). First of all, we can see that the value of $g^{(2)}(0)$ is inversely proportional to the electron beam current, which is in agreement with previous experimental results[1,2,6]. This can now be understood from the fact that the bunching peak scales linearly with the number of electrons reaching the sample (Eq. (3)), as it depends on the number of correlations between photons from the same electron. Instead, the background scales quadratically (Eq. (4)), since it depends on the events created by photons from different electrons.

Secondly, Eq. (5) shows that $g^{(2)}(0)$ does not depend on the number of carriers generated per bulk plasmon $m$, nor on the emission efficiency $\eta$ of these carriers. The only relevant parameter is the number of bulk plasmons created close to the QWs ($b$). This is in agreement with the fact that $g^{(2)}(\tau)$ measurements are independent of the absolute intensity incident on the detectors, as long as the statistics of the emission process is preserved[4]. Notice that even in the case $b = 1$, i.e. on average one bulk plasmon per electron interacts with the QWs, $g^{(2)}(0) > 1$ due to the stochastic nature of the plasmon excitation process. It follows from Eq. (5) that the bunching contribution to $g^{(2)}(0)$ increases with decreasing $b$, given that decreasing the number of interacting plasmons generated per electron would have a similar effect as decreasing the current. Given the Poissonian nature of the $b_i$ parameter, it is related to the probability of creating at least one interaction (bulk plasmon) close to the emitter[6], defined as

$$\gamma = 1 - \text{Poiss}(0; b) = 1 - e^{-b}, \tag{6}$$

where $\gamma$ can be interpreted as the excitation efficiency of the electron in the given material geometry.

Finally, the value of $g^{(2)}(0)$ also depends on shape of the bunching curve ($h_b(\tau)$), which is represented by the dependence on $\alpha_b$. Given an emitter decay $y_{\text{emitter}}(t)$, it can be shown that the number of correlations between photons emitted with a delay $\tau$ is proportional to (see S4a)

$$h_b(\tau) = \int_0^\infty y_{\text{emitter}}(t) y_{\text{emitter}}(t + \tau)\, dt. \tag{7}$$

In the case that the emitter decays as a simple exponential with lifetime $\tau_{\text{emitter}}$, $h_b(\tau)$ is an exponential with $\tau_{\text{emitter}}$, and thus the decay of the $g^{(2)}(\tau)$ curve directly gives the emitter decay. In this case, the relation between the area and the height of the bunching peak is: $\alpha_b = 2\tau_{\text{emitter}}$ (see S2e). In the case of more complex decays mechanisms, we should apply Eq. (7) to extract the emission dynamics from the $g^{(2)}(\tau)$ measurement. Eqs. (5) and (7) can now be directly used to fit experimental data of $g^{(2)}(\tau)$ versus beam current, to determine $b$, and hence $\gamma$, thus providing essential information on the electron beam excitation efficiency in incoherent CL excitation.

Figure 2a shows a selection of $g^{(2)}(\tau)$ measurements of the QW sample at different electron currents, all obtained using a continuous 10 keV electron beam. The time binning in all measurements is set to $t_b = 512$ ps. At the lowest current (2.8 pA), $g^{(2)}(0) = 12.6$ is obtained, while the value of $g^{(2)}(0)$ strongly decreases for increasing current. The curves cannot be properly fitted with a simple exponential decay, probably due to multiple decay processes taking place simultaneously. Instead, the emitter decay ($y_{\text{emitter}}(t)$) can be described with a stretched exponential, with parameters $\tau_{\text{emitter}}$ representing the average emitter lifetime and $\beta_{\text{emitter}}$ giving the deviation from a pure exponential decay[31]. This is further confirmed by direct measurements of the decay statistics of the QWs (see S4c). In this case, the shape of $g^{(2)}(\tau)$ does not give directly the emitter decay properties, but we need to fit the data with Eq. (7), which can only be solved numerically. The solid lines in the figures correspond to the fits, from which we obtain an emitter lifetime of $\tau_{\text{emitter}} = 10.65 \pm 0.32$ ns and $\beta = 0.76 \pm 0.01$.

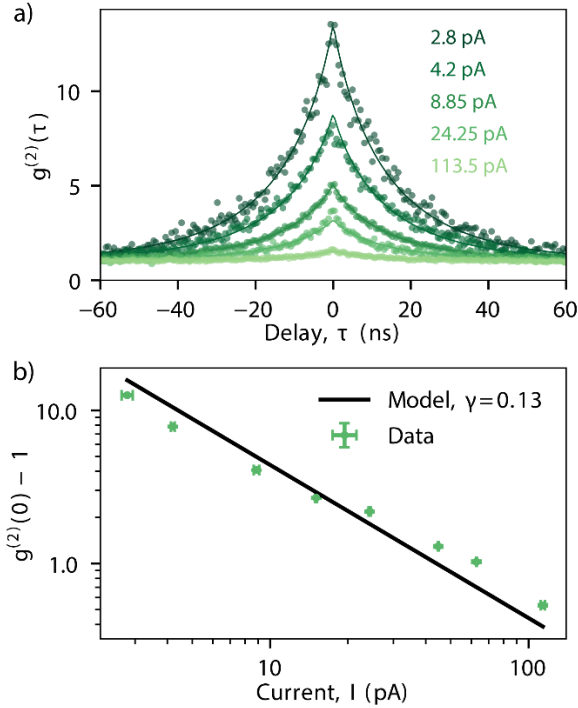

**Figure 2:** $g^{(2)}(\tau)$ measurements with a continuous e- beam. (a) $g^{(2)}(\tau)$ experiments obtained for different electron currents, all with an electron energy of 10 keV. The points represent the data and the solid lines the fit. (b) Fits of $g^{(2)}(0) - 1$ versus electron current obtained from the curves in (a). The black solid line is the fit obtained using Eq. (5), from which a value of $\gamma = 0.13$ is obtained. The error bars represent the uncertainty in the measured value of the electron current (horizontal) and fit errors (vertical).

In order to compare the experimental results with the analytical model, Fig. 2b shows the value of $g^{(2)}(0) - 1$, obtained from the fits of each curve, as a function of electron current. The horizontal error bars indicate the uncertainty in measuring the electron current, while the vertical errors are derived from the fitting errors. We observe that the value of $g^{(2)}(0) - 1$ exhibits a linear decrease (on a log-log plot) with slope $-1$, as predicted by Eq. (5). The shape factor $\alpha_b$ is calculated numerically from the solution of Eq. (7) using the derived values of $\tau_{\text{emitter}}$ and $\beta_{\text{emitter}}$, thus becoming $\alpha_b = 25.04$. Therefore, we can extract $b$ from Eq. (5). We obtain the best fit for $\gamma = 0.13 \pm 0.01$, meaning that on average, only 13 out of 100 electrons actually interact with the QWs. This low interaction between the electrons and the QWs is attributed to the fact that at 10 keV most electrons lose their energy before arriving to the QWs, as shown in Figure 1b and discussed in ref. [6]. Moreover, the carriers generated on the top GaN layer cannot reach the QWs, due to the presence of the AlGaN blocking layer on top of the QWs. For reference, we also show in the Supporting Information (section S1) that the results from the model are in excellent agreement with those obtained with the Monte Carlo-based approach proposed in previous works[1,6], confirming that our model serves as an effective analytical version of the Monte Carlo one.

*Pulsed electron beam*

$g^{(2)}(\tau)$ experiments can also be performed using pulsed electron beams, which can offer advantages such as lower acquisition times and simpler analysis, as will be discussed below. In this configuration, the photon emission dynamics is shaped by the temporal spread of electrons, and thus a modified model needs to be developed. A schematic of the histogram obtained in an HBT experiment in pulsed conditions is shown in Figure 1c (right). In contrast to the continuous case, here the histogram is composed of a peak at $\tau = 0$, containing correlations between photons from the same electron pulse, and peaks at delays corresponding to the time between pulses ($\tau_i = \frac{i}{F}$, with $i = \pm 1, \pm 2, \pm 3 \ldots$ being an integer number and $F$ the repetition rate). The latter correspond to correlations between photons from consecutive pulses ($i = \pm 1$), from every second pulse ($i = \pm 2$) and so on. These peaks are thus analogue to the background ($A_{\text{uncorr}}^c$) in the continuous case.

The derivation of the model for the pulsed case is similar to the one for the continuous one, with the main difference being the shape factor of the bunching ($\tau_0$) and uncorrelated ($\tau_i, i \neq 0$) peaks. Given that the peaks at $\tau_i$ ($i \neq 0$) contain correlations between photons from different pulses, their shape is determined by both the electron pulse and emitter decay, as explained in S4. The ratio between the area ($A_{\text{uncorr},i}^p$) and the height ($h_{\text{uncorr},i}$) of any of these peaks is given by $A_{\text{uncorr},i}^p = \alpha_{\text{conv}} h_{\text{uncorr},i}$. $\alpha_{\text{conv}}$ is thus a shape factor, which will depend on the particular shape of the electron pulse and emitter decay.

The peak centered at $\tau = 0$, accounts for correlations between photons from the same pulse, and has two components ($A_0^p = A_{\text{uncorr},0}^p + A_b^p$). The first component corresponds to correlations between photons from the same pulse, but different electrons, and therefore has a shape factor $\alpha_{\text{conv}}$. The second component corresponds to correlations between photons from the same electron, which is what constitutes the bunching. Similar to the continuous case, we can consider that all the excitations take place instantaneously, given that the timescale of bulk plasmon decay and carrier diffusion (typically in the fs/ps regime)[5] is much smaller than the emitter lifetime (hundreds of ps or ns). Therefore, the shape of the electron pulse does not play a role in this component, contrary to the case for the uncorrelated component. The time between photons is only determined by the emitter lifetime, and $A_b^p = \alpha_b h_b^p$, where $h_b^p$ is the height of this peak.

Taking into account the shape factors, and calculating the average number of possible combinations correlations for bunching ($A_b^p$) and uncorrelated ($A_{\text{uncorr},i}^p$) events in analogous way as in the continuous case (see

Supporting Information S3 for a full derivation), we obtain that for pulsed excitation

$$g^{(2)}_{\text{pulsed}}(0) = 1 + \frac{\alpha_{\text{conv}}}{\alpha_{\text{b}}} \frac{b+1}{n_e b}. \quad (8)$$

We observe that the expression for $g^{(2)}(0)$ for a pulsed beam is very similar to the one for the continuous case (Eq. (5)). Here, $g^{(2)}(0)$ is inversely proportional to the number of electrons per pulse, which is related to the electron beam current through $n_e = I/qF$, with $F$ being the repetition rate. The dependence of $g^{(2)}(0)$ on the average number of bulk plasmons that interact with the sample, $b$, is exactly the same as for a continuous electron beam, showing that the pulsed $g^{(2)}(\tau)$ measurement fundamentally probes the interaction of electrons with the sample in the same way. The main difference between the continuous and pulsed case is the factor $\alpha_{\text{conv}}$: in the pulsed case the $g^{(2)}(0)$ depends also on the shape of the electron pulse. From the derivation of $g^{(2)}(0)$ (section S3) it also follows that now we can simply divide the area of the peak at $\tau = 0$ by the area of any other peak at $\tau \neq 0$ to obtain the excitation efficiency:

$$\frac{A^p_b + A^p_{\text{uncorr},0}}{A^p_{\text{uncorr},i}} = 1 + \frac{b+1}{n_e b} = 1 + \frac{1 - \log(1-\gamma)}{n_e \log(1-\gamma)}. \quad (9)$$

In this case we do not need any fitting procedure nor prior knowledge of electron pulse shape or emitter decay, thus making the analysis even simpler. This becomes particularly useful when having small signal-to-noise ratios or non-trivial emitter decays or electron pulse shapes, in which cases fitting becomes challenging.

In order to test the model for the pulsed case, we performed experiments using an ultrafast beam blanker, in which a set of electrostatic plates is inserted inside the electron column. One of the plates is driven using a pulse generator, which is set to send a square signal with peak-to-peak voltage of 5 V and offset 2.5 V, while the other plate is grounded. This configuration allows us to obtain effectively square electron pulses with pulse width ($\Delta p$) determined by the repetition rate $F$ and duty cycle $D$. A characterization of the electron pulses is shown in the Supporting Information (S6a). In our experiments we kept the duty cycle fixed at $D = 95\%$ and varied the repetition rate from 0.2 to 6 MHz, resulting in pulse widths ranging from 200 ns down to 6 ns. Notice that an even smaller pulse width, down to 30 ps, can be obtained using the same ultrafast blanker in a different configuration[19], but long pulse widths were chosen to show the effect on the bunching peak. The current in continuous mode (i.e., without blanking) was kept constant ($I = 214$ pA) for all experiments, therefore changing the repetition rate results in a varying number of electrons per pulse, i.e.,

$$n_e = \frac{I}{q} \frac{1 - D/100}{F} = \frac{I}{q} \Delta p. \quad (10)$$

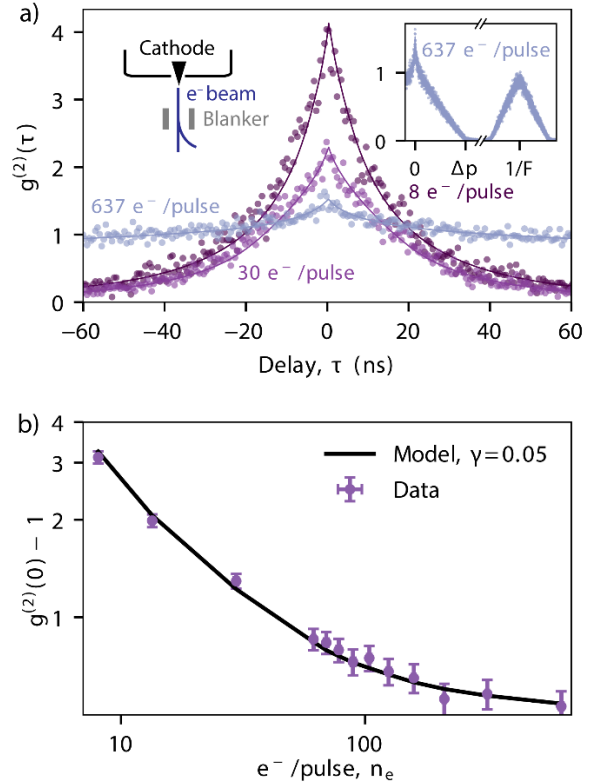

**Figure 3:** $g^{(2)}(\tau)$ measurements with ns-pulsed e⁻ beam. (a) $g^{(2)}(\tau)$ experiments obtained by changing the electron pulse width $\Delta p$, which leads to a change in the number of electrons per pulse. In this case the electron energy is 8 keV. The points represent the data and the solid lines the fit. Insets: (left) schematic of the beam blanking configuration, (right) example of a $g^{(2)}(\tau)$ measurement shown for a wider delay time range, thus showing the full peak at $\tau = 0$ and the consecutive peak at $\tau = 1/F$, where $F$ is the repetition rate. (b) Experimental results of $g^{(2)}(0) - 1$ vs. number of electrons per pulse, obtained from the fits of the curves in (a). The black solid line corresponds to the best fit obtained using Eq. (10), which yields an excitation efficiency of $\gamma = 0.05$. The error bars are derived from the uncertainty in the current measurement (horizontal) and fitting procedure (vertical).

Figure 3a shows a selection of $g^{(2)}(\tau)$ curves centered around the peak at delay 0. The solid lines are the fits of the data, which correspond to the sum of the solution from Eq. (7) (assuming a decay following a stretched exponential), and a convolution between a triangular curve and the same solution from Eq. (7). The triangular function comes from the convolution between two square pulses with pulse width ($\Delta p$), representing two electron pulses (see S4). The best fit of the curves is found for $\tau_{\text{emitter}} = 5.40 \pm 0.33$ ns and $\beta_{\text{emitter}} = 0.56 \pm 0.01$. The difference between these values and the ones found in the continuous experiment (10.7 ns and 0.67, respectively) is attributed to the inhomogeneity of the sample, which results in emission lifetimes that depend on sample position (see S8). The discrepancy could also come from the fact that at 10 keV we might be probing deeper QWs, which can exhibit different lifetimes. The curve at the lowest number of electrons per pulse (8 e−/pulse) exhibits the highest amplitude ($g^{(2)}(0) = 4.1$). In this case, the pulse width (6 ns) is comparable to the emitter lifetime, and thus no clear distinction between the bunching ($A_b^p$) and uncorrelated ($A_{\text{uncorr},0}^p$) curves can be observed. Instead, the $g^{(2)}(\tau)$ curve for long pulses show a small sharp peak, corresponding to the bunching peak, on top of a broader background, as can be observed in the curve corresponding to $\Delta p = 500$ ns (637 electrons per pulse). The full shape of the uncorrelated peak can be observed in the right inset of Figure 3a, showing the peak around delay 0 and the first consecutive peak ($\tau_1 = 1/F$).

The value of $g^{(2)}(0) - 1$ as a function of the number of electrons per pulse is shown in Figure 3b, which has been derived from the fits in Fig. 3a. We observe that the bunching decreases with increasing number of electrons per pulse, as expected, but, contrary to what we observed in the continuous case, the data does not exhibit a linear trend (on the log-log plot). This is due to the fact that in this comparison we are changing $\alpha_{conv}$ in each measurement. For a fixed beam current, large pulse widths correspond to a higher number of electrons per pulse. So, while we expect decreasing $g^{(2)}(0)$ with electrons per pulse, the factor $\alpha_{conv}$ also becomes larger, thus effectively increasing $g^{(2)}(0)$. The model, which accounts for this effect, shows a good agreement with the data. We can therefore extract an excitation efficiency of $\gamma = 0.05$. The fact that a lower $\gamma$ is found here compared to the continuous case is fully consistent with the fact that the pulsed experiments were performed with an electron energy of 8 keV instead of 10 keV. This choice of lower electron energy allows us to achieve relatively high $g^{(2)}(0)$ amplitudes despite having a high current on the sample (214 pA), due to the blanking conditions. At this lower electron energy, most bulk plasmons are generated in the top GaN layer, resulting in fewer excitations close to the quantum wells. The spectrum of the QWs and Casino simulations at 8 keV is provided in the Supporting Information (section S7). Additionally, we can derive the excitation efficiency using Eq. (10) by simply dividing the area of the bunching peak by the area of any other peak, from which we obtain $\gamma = 0.06$, which is in good agreement with the value found using the fitting procedure.

*Ultrashort pulses*

An extreme case of the model for pulsed $g^{(2)}(\tau)$ measurements is when we have ultrashort electron pulses, i.e., in the picosecond regime. In this case, the electron pulse width is very small compared to the emitter lifetime, and thus the factor accounting for the convolution of both becomes $\alpha_{conv} = \alpha_b$. Then, Eq. (10) can be further simplified to

$$g^{(2)}_{\text{ultrashort}}(0) = 1 + \frac{b+1}{n_e b} = \frac{A_0^p}{A_{\text{uncorr},i}^p}, \quad (11)$$

where the only remaining parameters are the number of interacting bulk plasmons per electron, which can be also described in terms of excitation efficiency $\gamma$, and the average number of electrons per pulse $n_e$. In this case the shape of the bunching peak, and thus, the emitter lifetime, do not contribute to the amplitude of $g^{(2)}(0)$. Moreover, $g^{(2)}(0)$ now can be directly obtained from the ratio between the areas of the different peaks, similar to Eq.(10). Hence, when analyzing an experiment, we can simply sum all the counts from each of these two peaks (at $\tau = 0$ and $\tau_i = \frac{i}{F}, i = \pm 1, \pm 2, \ldots$) and divide them to directly obtain the value of $g^{(2)}(0)$. In this way, the analysis to retrieve the excitation efficiency from $g^{(2)}(\tau)$ measurements becomes even simpler. We should also notice that ultrashort pulses are typically achieved by changing the emission statistics of the electron. For example, in the case of photoemission of electron pulses, as in the experiments shown below, the emission of pulses is determined by laser excitation of the electron cathode, instead of conventional thermoionic or Schottky emission. Our derivation of $g^2(0)$ is general and does not assume any particular emission statistics for the electron beam. In the Supporting Information (section S3e) we show a complementary derivation for electron pulses obtained with photoemission.

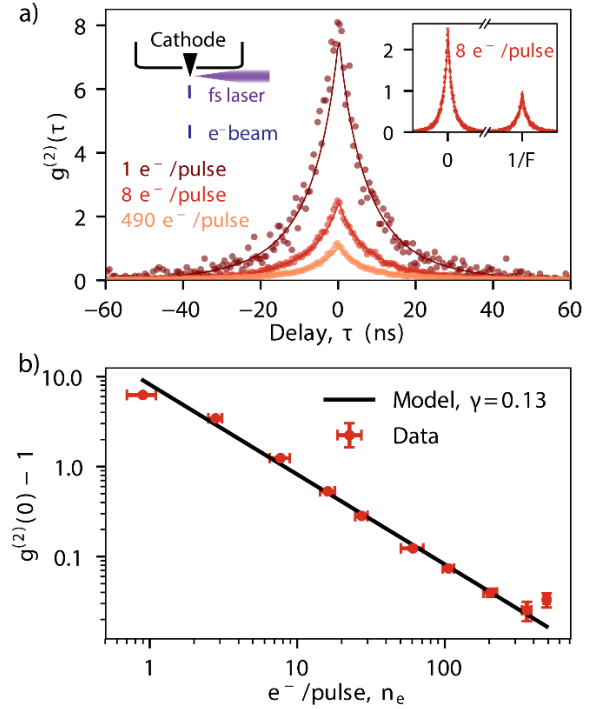

**Figure 4:** $g^{(2)}(\tau)$ measurements with ultrashort (ps) electron pulses. (a) $g^{(2)}(\tau)$ experiments obtained by changing number of electrons per pulse. The experimental data is represented by points, while the solid lines are the fits obtained by solving Eq. (7) (with $y_{\text{emitter}}(\tau)$ being a stretched exponential). Insets: (left) schematic of the photoemission setup, (right) zoom-out of a $g^{(2)}(\tau)$ measurement, showing that the shape of the uncorrelated peaks (in this case, $\tau = 1/F = 200\ ns$) is now determined only by the emitter decay. (b) $g^{(2)}(0) - 1$ vs. number of electrons per pulse, obtained by dividing the area of the bunching peak by the area of any other peak, as discussed in Eq. (13). The black solid line corresponds to the best fit obtained using Eq. (13), which yields an excitation efficiency of $\gamma = 0.13$.

Figure 4a shows a selection of $g^{(2)}(\tau)$ measurements performed using ultrashort pulses (~ 1 ps), obtained by focusing a fs laser into the electron cathode at a repetition rate of 5.04 MHz[8,20]. We chose the conditions for which a larger number of electrons per pulse can be achieved (up to 490 in this case) at the expense of spatial resolution[19]. This regime allows us to reach the highest possible electron cascade density, as will be discussed below. The experiments were performed with an electron energy of 10 keV. The figure shows the 0-delay peak for a changing number of electrons per pulse. We observe that with an average of 1 electron per pulse we obtain $g^{(2)}(0) = 7.25$. The right inset in Figure 4a shows a measurement including also the first uncorrelated peak, centered at $\tau = 198$ ns. We observe that now both the peaks at $\tau = 0$ and at $\tau = 1/F$ have the same shape, determined by the emitter decay. The solid lines again represent the curves obtained by fitting with Eq. (7), given that the emitter decay follows a stretched exponential. The best fit is obtained for $\tau_{emitter} = 5.84 \pm 0.07$ ns and $\beta_{emitter} = 0.751 \pm 0.004$, Figure 4b shows the value of $g^{(2)}(0) - 1$ as a function of the number of electrons per pulse, together with the fit using Eq. (13). Here, the data points have been obtained by simply dividing the areas of the bunching peak by the height of peaks at $\tau_i = i/F$. The horizontal error bars represent the uncertainty in current measurement in pulsed, partially due to instability in the power of the laser that excites the tip. The vertical error bars are obtained from the analysis of areas below the peaks. We also correct for the fact that the number of events decreases at long delays due to an experimental artifact (see S5). We observe that the data follows the trend predicted by Eq. (13), yielding the best fit for the model for $\gamma = 0.130 \pm 0.001$, which is in agreement with the excitation efficiency found in the experiments in continuous mode, in which the same electron energy was used. This confirms the feasibility of using the $g^{(2)}(\tau)$ analysis with ultrashort electron pulses to obtain the excitation efficiency, thus enabling many applications of $g^{(2)}(\tau)$ spectroscopy in ultrafast electron microscopy.

Even though the data show a linear trend as in the continuous case, we should note that the electron excitation is very different between the continuous and pulsed cases. In ultrafast pulsed mode, we are exciting the sample with a large number of electrons within a very short time (ps), while in the continuous or beam-blanked cases the average time between two consecutive electrons was never smaller than hundreds of ps (600 ps at 260 pA). We expect that bulk plasmons decay within the first tens of fs after electron excitation, initially creating hot carriers. The thermalization of these carriers typically occurs within tens of ps [5]. Therefore, in the ps-pulsed $g^{(2)}(\tau)$ experiment up to 490 electrons in each pulse excite the sample within the carrier thermalization time, and in a relatively small area. This raises the question whether we are inducing any nonlinear interaction between carriers due, such as Auger recombination, to high carrier concentrations.

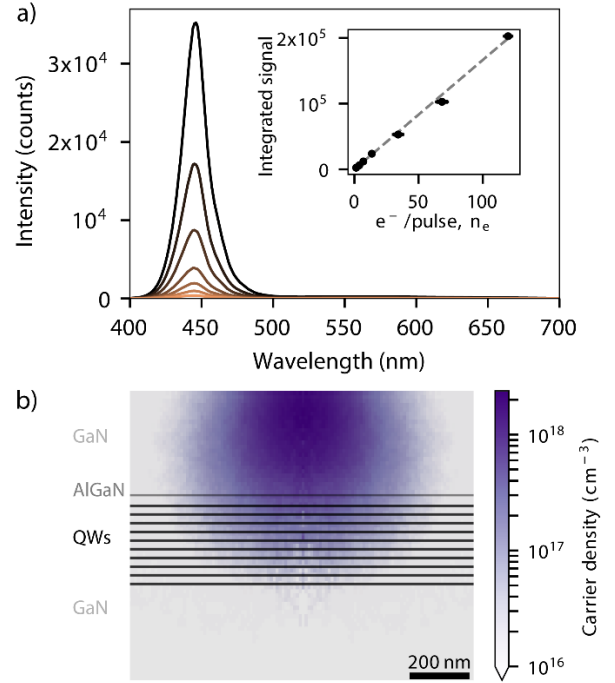

**Figure 5:** (a) Spectra of the QWs obtained when exciting with ultrashort electron pulses, containing from 1.4 (orange) up to 120 (black) e⁻ per pulse on average. Inset: integrated signal of the QW emission as a function of the number of electrons per pulse. The dashed line corresponds to a linear fit. (b) Simulated 2D map of the carrier density in the sample obtained after excitation of 490 electrons with an energy of 10 keV.

Previous work on InGaN/GaN quantum wells under optical excitation showed that a high excitation fluence leads to a decrease in efficiency, typically referred to as 'efficiency droop'[32]. Even though the origin of this effect is still under discussion, some works attributed this efficiency droop to Auger recombination due to locally-induced high carrier densities[32]. Nevertheless, the trend of $g^{(2)}(0)$ with the number of electrons per pulse exhibits a clear power law, as expected from the $g^{(2)}(\tau)$ model, which neglects nonlinear effects. These results suggest that nonlinear interactions between carriers do not play a significant role in this case, even at the highest number of electrons per pulse. This further suggests that the induced carrier densities are lower than the threshold for Auger recombination to occur.

To further elucidate this absence of nonlinear effects, Figure 5a shows CL spectra of the QWs obtained at different number of electrons per pulse. The integrated area below the curve (in the 410-490 nm spectral range) as a function of the number of electrons per pulse is plotted in the inset of Figure 5a. We observe a clear linear trend with increasing number of electrons per pulse. Figure 5b shows the calculated carrier density for a 10 keV electron beam containing 490 electrons. Here we have assumed a radius of the electron beam of 200 nm, corresponding to the expected spatial resolution obtained under our pulsed conditions, calculated using the Fourier transform method explained in ref. [19]. We use Casino simulations to calculate the number of inelastic collisions of the electron with the sample. We assume that each collision corresponds to the generation of a bulk plasmon and generates 3 electron/hole pairs[33,34]. We observe that the highest carrier density is in the order of $6 \times 10^{17}$ cm$^{-3}$. Previous works based on optical excitation of InGaN show that Auger processes only become dominant for carrier densities larger than $10^{18}$ cm$^{-3}$[35,36]. Therefore, the electron-induced carrier density is below that which would create nonlinear effects. Moreover, we expect the initial spatial distribution of carriers to be relatively localized in space after electron excitation, implying that diffusion of carriers plays a larger role than in optical experiments, in which the spot size is limited by diffraction. We note that this is the largest number of electrons per pulse that we can obtain in our system at 10 keV. Other pulsed conditions lead to better spatial resolution and hence more confined electron cascades, but at much lower number of electrons per pulse (less than tens of e⁻ per pulse)[19]. Other works have shown a spatial resolution in the nm range, but in the regime of few electrons per pulse, and thus small electron density[37,38].

*Conclusions*

In conclusion, we have presented a full description of $g^{(2)}(\tau)$ autocorrelation measurements in incoherent cathodoluminescence spectroscopy for different electron beam configurations. We have developed a fully analytical model to explain the amplitude of bunching ($g^{(2)}(0)$) as a function of electron beam current (or number of electrons per pulse), electron excitation efficiency, emitter lifetime and pulse duration, in the case of pulsed electron beams. The model highlights the inverse dependence of the bunching contribution to $g^{(2)}(\tau)$ as a function of electron beam current or number of electrons per pulse. Moreover, by acquiring a $g^{(2)}(\tau)$ curve at a known electron beam current we can directly extract the electron excitation efficiency by using a simple equation, and the curve can be fitted to obtain the emitter lifetime. This is a major step forward compared to the previous method in which Monte Carlo simulations were needed, given the simplicity of the analysis using our model.

In particular, we show that for a pulsed electron beam, the excitation efficiency can be obtained by simply dividing the areas of the peak at 0 delay by that of any other peak, without the need of fitting the data. The model is generic and independent of the sample under study and prior knowledge of the sample geometry. In order to test the model with experiments, we have studied InGaN/GaN quantum well samples, in which we find an excitation efficiency of 0.13 for 10 keV electrons and 0.05 in the case of 8 keV electrons. Furthermore, we have presented $g^{(2)}(\tau)$ CL measurement using ultrashort (ps) electron pulses, with the average number of electrons ranging from less than 1 to ~500. The measurements of $g^{(2)}(0)$ as a function of the number of electrons per pulse exhibit the same trend as predicted by the analytical model, suggesting that nonlinear carrier interactions do not play a role, even at a high number of electrons per pulse. We model the induced carrier density in the QW sample and show that it remains lower than typical values for which nonlinear effects in optical excitation are observed. We foresee that the analytical model will make $g^{(2)}(\tau)$ measurements and analysis more accessible, thus allowing to get deeper insights into the fundamentals of electron-matter interaction. Moreover, the $g^{(2)}(\tau)$ experiments with ultrashort pulses pave the way to study photon statistics with dense electron cascades in a wide range of materials.

## AUTHOR INFORMATION

### Corresponding Author

* Email: m.sola@amolf.nl

### Notes

The authors declare the following competing financial interests: T. C is employee and A.P. is co-founder and co-owner of Delmic BV, a company that produces the cathodoluminescence system that was used in this work.

## ACKNOWLEDGMENT

The authors would like to thank Silke Christiansen and Michael Lätzel for providing the sample. This work is part of the research program of AMOLF which is partly financed by the Dutch Research Council (NWO). This project has received funding from the European Research Council (ERC) under the European Union's Horizon 2020 research and innovation program (grant agreement No. 695343). S.M. acknowledges support from the French ANR funding agency through the ANR-19-CE30-0008-ECHOMELO grant.

# Supplementary information of

# Photon statistics of incoherent cathodoluminescence with continuous and pulsed electron beams

Magdalena Solà-Garcia,[1,*] Kelly W. Mauser [1], Matthias Liebtrau, [1] Toon Coenen, [1,2] Sophie Meuret, [3] Albert Polman[1]

[1] Center for Nanophotonics, AMOLF, Science Park 104, 1098 XG Amsterdam, The Netherlands
[2] Delmic BV, Kanaalweg 4, 2628 EB, Delft, The Netherlands
[3] CEMES-CNRS, 29 Rue Jeanne Marvig, 31055 Toulouse, France

[*] Corresponding author: m.sola@amolf.nl

**Table of contents:**

# S1. Comparison of analytical model to Monte Carlo simulations

Previous $g^{(2)}(\tau)$ measurements in cathodoluminescence (CL) have been modelled using Monte Carlo (MC) simulations. Here we demonstrate the accuracy of our analytical model by comparing $g^{(2)}(0)$ results obtained with our analytical model to those produced by MC simulations. The comparison is performed for the three different electron beam configurations (continuous, pulsed with beam blanker and pulsed through photoemission). In all cases, the following steps were considered after the arrival of an electron to the sample:

1. Creation of $b_i$ bulk plasmons, according to a Poisson distribution with expectation value $b$.
2. Decay of each plasmon into $m_i$ electron-hole pairs, described with a Poisson distribution with expectation value $m$.
3. Excitation of a quantum well by an electron-hole pair with probability $\eta$.
4. Emission of a photon, following a given decay mechanism.

The MC simulations with a continuous electron beam have been performed using the same code as in refs. [1-3]. The code was adapted to represent the blanker experiment, in which only part of the initial continuous beam reaches the sample, thus generating (relatively long) electron pulses. In the MC simulations for the blanker case the current in continuous mode and the repetition rate were set to 20 pA and 1 MHz, respectively, and the pulse width was varied from 8 up to 500 ns, similar to the experiments. We also adapted the initial MC code to simulate electron pulses generated by photoemission. In this case, no continuous electron beam is initially generated, but instead we directly create pulses containing a certain number of electrons per pulse, given by a Poisson distribution with expectation value $n_e$. The pulse width is assumed to be Gaussian, with $\sigma = 1$ ps. The exact value of the pulse width is not critical, given that it is much shorter than the lifetime $\tau_{\text{emitter}}$. The MC simulations with photoemission were performed assuming a repetition rate of $F = 5.04$ MHz. In all cases, the results from the simulations have been analyzed using the same procedure as for the experimental data.

Figure S1 shows the values of $g^{(2)}(0) - 1$ obtained from the MC simulations using a continuous electron beam (a), and a pulsed electron beam generated by beam blanking (b) and photoemission (c). In the three cases we show $g^{(2)}(0) - 1$ as a function of electron beam current (a) and number of electrons per pulse (b, c). In all cases we consider an exponential decay for the emitter, with lifetime $\tau_{\text{emitter}} = 12$ ns and an average number of $b = 0.2$ bulk plasmons per electron that interact with the quantum wells, corresponding to an excitation efficiency of $\gamma = 0.18$. We also assume $m = 1$ and $\eta = 1$, even though it has already been shown that these parameters do not play a role in the final result of the MC simulation[2]. In this work we explain this fact by showing that $m$ and $\eta$ cancel out in the development of the analytical model. The time step in the simulations was set to 512 ps, the same as in our experiments.

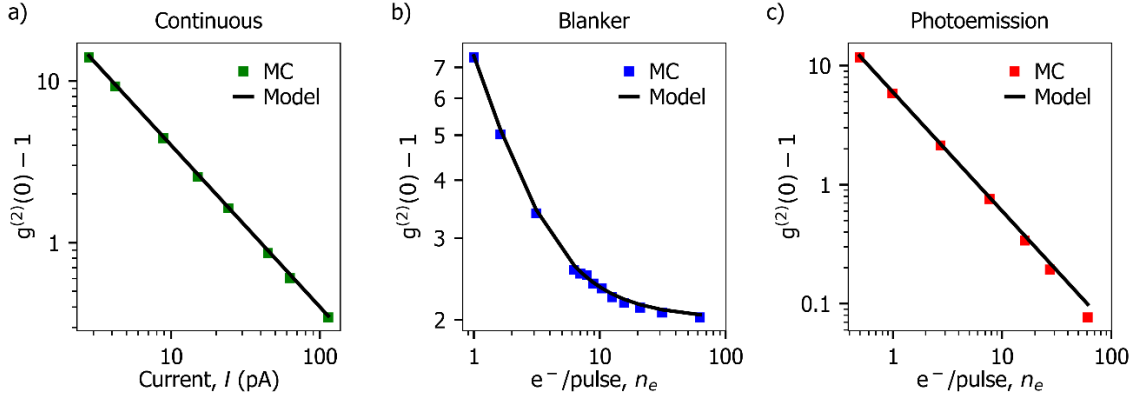

**Figure S1:** Monte Carlo simulations of $g^{(2)}(0)$ amplitude vs. electron beam current or number of electrons per pulse, together with the results from the analytical model. (a) Continuous electron beam, (b) electron beam pulsed using the blanking technique and (c) pulsed electron beam generated by photoemission.

Figure S1 also shows the results of our analytical model, in which we used the same parameters as in the MC simulations. We should note that here no fitting is needed, given that we just fix all the parameters (including $b$). The results show a very good agreement between the MC simulations and the analytical model.

## S2. Analytical model – Continuous electron beam

A $g^{(2)}(\tau)$ experiment measures the photon statistics of a given emitter or source, and it is therefore it is based on random processes: the emission of a photon is stochastic, following a certain probability distribution (for example, an exponential decay). Moreover, in typical experimental setups (such as the HBT experiment), the emitted photons are split randomly towards the two detectors, with a 50% probability of being detected by each detector[4]. The measurement is based on collecting enough statistics such that it can accurately represent the chances of having a correlation at a given delay compared to any other. It is therefore not possible to predict exactly how many photons will correlate with photons from the same electron (thus leading to bunching), and how many with photons from other electrons (uncorrelated events). Instead, we can calculate how likely it is that one scenario happens with respect to the other one. Hence, our analytical model is based on calculating the average number of combinations of correlations that lead to bunching ($A_\text{b}$) compared to the average number of combinations that lead to uncorrelated events (photons coming from different electrons, or pulses, in the pulsed case, $A_\text{uncorr}$).

We start from Eq. (2) of the main text, in which:

$$g^{(2)}_{cont}(0) = 1 + \frac{A_b^c}{A_\text{uncorr}^c} \frac{(2B+1)t_B}{2\alpha_\text{b}}, \quad \text{(S1)}$$

where $A_b^c$ contains the mean number of combinations of correlations between photons from the same electron (i.e., giving bunching), $A_\text{uncorr}^c$ is the mean number of combinations of correlations between photons from different electrons (uncorrelated), $B$ is the total number of bins during the acquisition time $T$, $t_B$ is the bin size and $\alpha_b$ is the shape factor of the bunching peak, defined as the ratio between the area and height of the peak. Hence, we now need to calculate $A_\text{b}^c$ and $A_\text{uncorr}^c$.

## S2a. Correlations between photons from the same electron ($A_b^c$, bunching peak)

We consider that when one electron interacts with an emitter (semiconductor, quantum wells, atomic defect, etc), a certain number of photons $k$ will be emitted, each of them with a certain arrival time $t_k$. We should note that this value $k$ is not fixed, but will be different for each electron, given the stochastic nature of all the processes (creation of bulk plasmon, decay into electron-hole pairs, radiative decay of emitter). We want to count the number of combinations of correlations between photons from the same electron. We define one correlation as the detection of a pair of photons, thus we need to take subsets of 2 from $k$ photons, each photon with a fixed arrival time. Moreover, the order matters, given that this will determine whether the measured delay between photons is positive or negative, and there are no repetitions, i.e., a photon cannot correlate with itself. This is a common problem in combinatorics[5], sometimes referred to as variation without repetition, from which we extract that the number of possible combinations is

$$A_k = \binom{k}{2} = k(k-1). \tag{S2}$$

Next, we want to relate $A_k$ to physical variables, i.e., expected value of number of bulk plasmons per electron ($b$), expected value of number electron-hole pairs created per plasmon ($m$) and radiative decay efficiency ($\eta$). We will follow steps 1-3 described in the main text, starting from step 3 and building up.

3. Given $m_i$ electron-hole pairs, each of them with a probability $\eta$ of exciting a QW that emits a photon, the expected value of the number of possible combinations of correlations of photons becomes

$$A_3 = \sum_{k=0}^{m_i} A_k \mathrm{Bin}(k; m_i, \eta) = \sum_{k=0}^{m_i} k(k-1) \frac{m_i!}{k!(m_i-k)!} \eta^k (1-\eta)^{m_i-k} = \eta^2 m_i (m_i - 1). \tag{S3}$$

2. Each bulk plasmon will create $m_i$ electron-hole pairs, described with a Poisson distribution with expected value $m$ $\left(\mathrm{Poiss}(m_i; m) = \frac{e^{-m} m^{m_i}}{m_i!}\right)$. Hence, we need to account for all the possible values of $m_i$, weighted by their probability. The expected value of the number of possible combinations correlations of photons produced by $m_i$ electron-hole pairs is then

$$A_{2,b_i=1} = \sum_{m_i=0}^{\infty} N_3 \mathrm{Poiss}(m_i; m) = \sum_{m_i=0}^{\infty} \eta^2 m_i (m_i - 1) \frac{e^{-m} m^{m_i}}{m_i!} = m^2 \eta^2. \tag{S4}$$

If an electron creates more than one bulk plasmon, each of these plasmons will decay into a certain amount of electron-hole pairs, with likelihood given by a Poisson distribution with expectation value $m$, as already described. Thus, we need to account for all the possible combinations of correlations of photons produced by an arbitrary number $b_i$ of bulk plasmons. We start with the case of two bulk plasmons, in which the expected value of the number of possible combinations of correlations of photons (the correlations can be from photons from the same or different plasmon) becomes

$$A_{2,b_i=2} = \sum_{m_1=0}^{\infty} \sum_{m_2=0}^{\infty} \eta^2 (m_1 + m_2)(m_1 + m_2 - 1) \, \mathrm{Poiss}(m_1; m) \mathrm{Poiss}(m_2; m) = 4m^2 \eta^2. \tag{S5}$$

It can be shown by induction (see S2d) that in the general case of $b_i$ bulk plasmons, which produce photons that can correlate with other photons from the same plasmon or a different plasmon, the expectation value of the number of possible combinations of correlations is

$$A_{2,b_i} = b_i^2 m^2 \eta^2. \tag{S6}$$

1. Finally, the number of bulk plasmons produced by a single electron also follows a Poisson distribution with expected value $b$ (Poiss($b_i; b$)). Therefore, averaging again over all possible values of $b_i$, we obtain that the average number of possible combinations of correlations produced by one electron is

$$A_1 = \sum_{b_i=0}^{\infty} b_i^2 m^2 \eta^2 \text{Poiss}(b_i; b) = bm^2\eta^2(b+1). \tag{S7}$$

In the case of $n$ electrons, the mean number of possible combinations of correlations between photons from the same electron becomes:

$$A_b^c = nA_1 = nbm^2\eta^2(b+1). \tag{S8}$$

### S2b. Correlations between photons from different electrons ($A_1^c$)

Next, we need to calculate the number of possible combinations of correlations between photons from different electrons. Taking into account the statistical distributions of each parameter involved in the emission of a photon (bulk plasmons, carriers, emission efficiency), the average number of photons emitted per electron is

$$N_{\text{ph}} = \sum_{b_i=0}^{\infty} \sum_{m_i=0}^{\infty} \sum_{k=0}^{m_i} k b_i \text{Poiss}(b_i; b) \, \text{Poiss}(m_i; m) \, \text{Bin}(k; m_i, \eta) = bm\eta. \tag{S9}$$

We should note that the result is the same as if we would just consider the expected values $b$, $m$ and $\eta$ given that the number of photons scales linearly with these parameters. We now need to create pairs between two photons from different electrons. In this case the order is still important. We calculate the average number of combinations of correlations of photons coming from different electrons as

$$A_{\text{uncorr}}^c = [nbm\eta][(n-1)bm\eta] = n(n-1)b^2m^2\eta^2. \tag{S10}$$

### S2c. $g^{(2)}(0)$ for a continuous beam

Finally, we can insert Eqs. (S8) and (S10) into Eq. (S1), rewrite $n$ as a function of the electron current, $I = nq/(Bt_b)$, and take the limit $B \to \infty$ to obtain

$$g_{\text{cont}}^{(2)}(0) = \lim_{B \to \infty} \left(1 + \frac{(2B+1)t_b}{2\alpha_b} \frac{b+1}{\left(I/q \, t_b B - 1\right)b}\right) = 1 + \frac{q}{I\alpha_b} \frac{b+1}{b}. \tag{S11}$$

The last expression can also be expressed in terms of the excitation efficiency γ (Eq. (6) in the main text, further explained in section S2e) such that

$$g_{\text{cont}}^{(2)}(0) = 1 + \frac{q}{I\alpha_b} \frac{\log(\gamma-1)-1}{\log(\gamma-1)}. \tag{S12}$$

## S2d. Bunching peak: mean number of possible combinations of photon correlations from $b_i$ bulk plasmons

We want to find the expected value for number of combinations of correlations for an arbitrary number of bulk plasmons. Similar to Eq. (S5), in the case of $j + 1$ bulk plasmons, we have

$$A_{2,j+1} = \eta^2 \sum_{m_1=0}^{\infty} \ldots \sum_{m_j=0}^{\infty} \sum_{m_{j+1}=0}^{\infty} (\bar{m} + m_{j+1})(\bar{m} + m_{j+1} - 1) \, P_{\bar{m}} P_{j+1}$$

$$= \eta^2 \sum_{m_1=0}^{\infty} \ldots \sum_{m_{b_j}=0}^{\infty} \sum_{m_{j+1}=0}^{\infty} (m_{j+1}^2 + m_{j+1}(2\bar{m} - 1) - \bar{m} + \bar{m}^2 ) \, P_{\bar{m}} P_{j+1},$$

(S13)

where we have defined $\bar{m} = m_1 + \ldots + m_j$ and $P_{\bar{m}}$ is the product of Poisson distributions, i.e., $P_{\bar{m}} = \text{Poiss}(m_1; m) \ldots \text{Poiss}(m_j; m) = \prod_{i=1}^{j} \frac{e^{-m} m^i}{i!}$. Eq. (S13) can be further developed into

$$A_{2,b=j+1} = \eta^2 \left[ m(m + 1) + 2jm^2 - m - jm + \sum_{m_1=0}^{\infty} \ldots \sum_{m_j=0}^{\infty} \bar{m}^2 P_{\bar{m}} \right].$$

(S14)

Therefore, we need to find an analytical expression for the last term in Eq. (S14). We assume that

$$\sum_{m_1=0}^{\infty} \ldots \sum_{m_j=0}^{\infty} \bar{m}^2 P_{\bar{m}} = jm(jm + 1),$$

(S15)

which we will prove by induction. In the case of $j = 1$,

$$\sum_{m_1=0}^{\infty} m_1^2 \, P(m_1; m) = m(m + 1).$$

(S16)

Assuming that Eq. (S15) is true, we need to prove that in the case of $j + 1$ bulk plasmons, it becomes $(j + 1)m[(j + 1)m]$. Hence

$$\sum_{m_1=0}^{\infty} \ldots \sum_{m_j=0}^{\infty} \sum_{m_{j+1}=0}^{\infty} (\bar{m} + m_{j+1})^2 P_{\bar{m}} P_{j+1} = \sum_{m_1=0}^{\infty} \ldots \sum_{m_j=0}^{\infty} \sum_{m_{j+1}=0}^{\infty} (\bar{m}^2 + 2\bar{m} m_{j+1} + m_{j+1}^2) P_{\bar{m}} P_{j+1}$$

$$= jm(jm + 1) + 2jm^2 + m(m + 1) = (j + 1)m[(j + 1)m].$$

(S17)

Finally, inserting Eq. (S15) into Eq. (S14), we obtain

$$A_{2,b=j+1} = (j + 1)^2 m^2 \eta^2.$$

(S18)

## S2e. Obtaining the excitation efficiency ($\gamma$)

For each electron, the probability of interacting, i.e., creating at least one plasmon that can excite the quantum wells, or any other emitter, is

$$P_{int} = 1 - \text{Poiss}(0; b) = 1 - e^{-b},$$

(S19)

where $b$ is the average number of bulk plasmons generated per electron (around the emitter). We define $\gamma$ as the fraction of electrons that create at least one bulk plasmon near the emitter. Given a certain number of electrons $n_{total}$, from which $n_{interacting}$ interact with the emitter, $\gamma$ becomes

$$\gamma = \frac{n_{interacting}}{n_{total}} = \frac{n_{total} P_{int}}{n_{total}} = 1 - e^{-b}. \tag{S20}$$

## S2f. Number of correlations at long delays

We consider that electrons interact with the sample during a certain (square) time window $T = Bt_b$, where $B$ is the total number of bins and $t_B$ is the bin size. The distribution of the electrons in time can be represented as a uniform random distribution. The number of possible correlations between photons coming from different electrons as a function of delay $\tau$ exhibits a triangular shape, with base corresponding to $2T$. This shape results from the convolution of two squared signals with width $T$. Thus, in the model, the total number of correlations are spread within an area corresponding to a triangle, with base $(2B+1)t_b$ and height $h_1$. Figure S2 shows an example of this effect. In the experiments, the typical acquisition time (at least seconds) is much larger than the time window within which we acquire correlations (30 $\mu s$ in our case for $t_b = 0.512$ ps), and thus this effect becomes negligible in the narrow time window in which we analyze the data, i.e., we only see the very top of the triangle. However, in the case of a pulsed electron beam, the time window corresponds to the pulse width $\Delta p$. The correlations between photons from the same or different pulses then exhibit a triangular shape, with base $2\Delta p$, as a function of $\tau$. This is the shape that we observe in our $g^{(2)}(\tau)$ measurements with the blanker (inset of Fig. 3, main text).

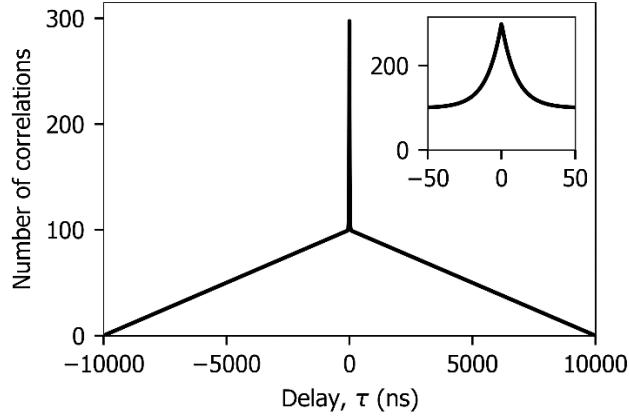

**Figure S2:** Theoretical shape of the number of possible correlations as a function of delay $\tau$ between photons emitted after the exposure of the sample to an electron beam during a time $T$. Here, $T = 10$ $\mu s$. The observed triangular shape results from the convolution of two square pulses with base $T$. The inset shows the curve at small delay, in which the triangular background is not observed.

## S2e. Calculation $\alpha_0$. Discussion discrete/continuous

A key parameter in our analytical model is the relation between the height and the area of the bunching peak, $\alpha_b$ (shape factor). Given a known decay function of the bunching peak, the shape factor can typically be easily calculated. In the case of a simple exponential we obtain $\alpha_b = 2\tau_b$, while for stretched exponential the shape factor becomes $\alpha_b = 2\frac{\tau}{\beta}\Gamma\left(\frac{1}{\beta}\right)$. Nevertheless, $g^{(2)}(\tau)$ measurements are discrete, and thus these expressions for $\alpha_b$ are only valid if the bin size $t_{bin}$ is much smaller than the typical decay time, such that we can assume an almost continuous function. Otherwise, the discretized nature of the measurement should be taken into account. For example, the generalized expression for $\alpha_{\text{bunching}}$ in the case of an exponential decay with arbitrary bin size is $\alpha_b = \frac{2t_b}{1-e^{-t_b/\tau_b}}$, which becomes $2\tau_b$ when $t_b \ll \tau_b$.

## S3. Analytical model – Pulsed electron beam

In the case of a pulsed electron beam we need to adapt the definition of $g^{(2)}(0)$ given in Eq. (1) of the main text. Here, we need to normalize the height at $\tau = 0$, with respect to the height of any other peak, $H_i$, which represents the uncorrelated events. Hence

$$g^{(2)}_{\text{pulsed}}(0) = \frac{H(0)}{H_i(i \neq 0)}. \tag{S21}$$

$H(0)$ is the height of the peak at $\tau = 0$. This peak will contain two contributions: correlations between photons from the same electron (with mean number of possible combinations $A_b^p$) and between photons from different electron but same pulse ($A_{\text{uncorr},0}^p$). As discussed in main text, the first contribution will be distributed over a temporal shape (with height $h_b^p$ and area $A_b^p$) determined by the emitter decay, through Eq. (S35). Hence, the ratio between the area and the height is: $h_b^p = \alpha_b A_b^p$, similar to the bunching contribution in the continuous case. In contrast, the temporal distribution of the correlations between photons from different electrons but same pulse (second contribution of $\tau = 0$ peak) depends not only on the emitter decay but also the shape of the electron pulse. Hence, the shape of this contribution is given by the convolution of two electron pulses, convoluted also with the emitter decay (see S4b). We define the ratio between the area ($A_{\text{uncorr},0}^p$) and the height ($h_{\text{uncorr},0}^p$) of this part as: $A_{\text{uncorr},0}^p = \alpha_{conv} h_{\text{uncorr},0}^p$.

$h_{\text{uncorr},i}(i \neq 0)$ is the height of any peak at $\tau \neq 0$, i.e., containing correlations between photons from consecutive pulses ($i = \pm 1$), from every second pulse ($i = \pm 1$) and so on. The shape of any of these peaks is also determined by the electron pulse shape and emitter decay, hence we can define: $A_{\text{uncorr},i}^p = \alpha_{conv} h_{\text{uncorr},i}^p$. $A_{\text{uncorr},i}^p$ contains the possible correlations between photons from different pulses.

Taking the previous definitions into account, we can rewrite (S21) as

$$g^{(2)}_{\text{pulsed}}(0) = \frac{h_b^p + h_{\text{uncorr},0}^p}{h_{\text{uncorr},i}^p} = \frac{\alpha_{conv} A_b^p + \alpha_b A_{\text{uncorr},0}^p}{\alpha_b A_{\text{uncorr},i}^p}. \tag{S22}$$

### S3a. Correlations between photons from the same pulse

#### Correlations between photons from the same electron ($A_b^p$)

The mean number of possible combinations of correlations between photons from the same electron, i.e., leading to bunching, is given in Eq. (S7), which we have to multiply by the number of electrons per pulse and the total number of pulses $r$

$$A_b^p = rb(b+1)m^2\eta^2 \sum_{n_i=0}^{\infty} n_i \text{Poiss}(n_i; n_e) = rn_e b(b+1)m^2\eta^2. \tag{S23}$$

Here we have assumed that the number of electrons per pulse $n_i$ follows a Poisson distribution with expected value $n_e$. This will be the case in most experiments, such as in the beam blanker and photoemission of electron pulses described in the main text. However, we would obtain the same result if we consider the number of electrons per pulse fixed, given that $A_b^p$ scales linearly with $n_i$.

## Correlations between photons from different electron within the same pulse ($A_{uncorr,0}^p$)

Given the average number of emitted photons per electron $N_{ph}$ (Eq. (S9)), the number of combinations of correlations between photons from the same pulse, but different electron, becomes

$$A_{uncorr,0}^p = b^2 m^2 \eta^2 \sum_{n_i=0}^{\infty} n_i(n_i - 1) \text{Poiss}(n_i; n_e) = r n_e^2 b^2 m^2 \eta^2. \tag{S24}$$

## S3b. Correlations between photons from different pulses ($A_{uncorr,i}^p$)

Following from Eq. (S9), which gives the average of photons emitted per electron, and assuming $n_i$ electrons per pulse (Poisson-distributed), the average number of photons emitted per pulse is

$$N_p = bm\eta \sum_{n_i=0}^{\infty} n_i \text{Poiss}(n_i; n_e) = n_e bm\eta. \tag{S25}$$

The number of possible correlations between photons from different pulses is therefore

$$A_{uncorr,r}^p = r(r-1)n_e^2 b^2 m^2 \eta^2, \tag{S26}$$

which is distributed over $2(r-1)$ peaks, given that we do not count the peak at $\tau = 0$, which would contain correlations between photons from the same pulse. We also need to take into account the peaks at $\tau_i$ are contained within a triangular envelope, given that the number of possible correlations decreases as the delay between pulses increases, as explained in section S2f.

Hence, the area below each peak at $\tau_i$ becomes

$$A_{uncorr,i}^p = \frac{2\, A_{uncorr,r}^p}{2(r-1)} = r n_e^2 b^2 m^2 \eta^2. \tag{S27}$$

## S3c. $g^{(2)}(0)$ for a pulsed electron beam

Finally, inserting Eqs. (S23), (S24) and (S27) into Eq. (S22) we obtain

$$g^{(2)}_{pulsed}(0) = 1 + \frac{\alpha_{conv}}{\alpha_b} \frac{b+1}{n_e b} = 1 + \frac{\alpha_{conv}}{n_e \alpha_b} \frac{\log(\gamma-1) - 1}{\log(\gamma-1)}. \tag{S28}$$

In which again we have used the relation between $b$ and $\gamma$ given in Eq. (6) of the main text.

## S3d. Alternative calculation of $\gamma$ for a pulsed electron beam

In the case of a pulsed electron beam, we don't need to calculate $g^{(2)}(0)$ to retrieve the excitation efficiency $\gamma$, but we can simply divide the sum of Eqs. (S23) and by Eq. (S27), which results in

$$\frac{A_b^p + A_{uncorr,0}^p}{A_{uncorr,i}^p} = 1 + \frac{b+1}{n_e b} = 1 + \frac{1 - \log(1-\gamma)}{n_e \log(1-\gamma)}, \tag{S29}$$

which corresponds to Eq. (10) in the main text. In experiments, this ratio would be equivalent to dividing the sums of all the counts below the peak at 0 delay with the sum of the counts below any other peak.

### S3e. Alternative derivation of $g^{(2)}_{\text{pulsed}}(0)$ in photoemission

The previous derivation of $g^{(2)}(0)$ assumes that bunching comes only from correlations between photons from the same electron. However, in the case of electron pulses obtained by photoemission, several electrons might excite the sample instantaneously (i.e., within a ps timescale, much smaller than the emitter decay). In this case bunching comes from correlations between photons from the same pulse, and it doesn't matter whether they come from the same or different electrons. Here we show a derivation of $g^{(2)}(0)$ starting from the point that all electrons within a pulse will create bunching, and show that it results in the same expression as Eq. (12) (main text).

We assume that the duration of the electron pulses is much smaller than the emitter lifetime. Hence, all peaks will have the same shape. In particular, the area below the bunching peak (peak at $\tau = 0$) is related to its height as: $A'_b = \alpha_b h'_b$. Similarly, any other peak $\tau_i (i \neq 0)$ follows the same relation: $A'_i = \alpha_i h'_i$. Eq. (S22) can now be written as

$$g^{(2)}_{\text{ultrashort,v2}}(0) = \frac{A'_b}{A'_i}. \tag{S30}$$

We first calculate the bunching peak, i.e., $A'_b$. The first steps are the same as in the continuous case. From Eq. (S6) we know that given $b_i$ bulk plasmons, the mean number of combinations of correlations is $b_i^2 m^2 \eta^2$. Assuming that we have $n_i$ electrons per pulse, each of them can create a different number of plasmons $b_i$. The case of $n_i = 1$ is derived in Eq. (S7). In the case of $n_i = 2$,

$$A_{n_i=2} = \sum_{b_1=0}^{\infty} \sum_{b_2=0}^{\infty} (b_1 + b_2)^2 m^2 \eta^2 \text{Poiss}(b_1; b) \text{Poiss}(b_2; b) = 2bm^2\eta^2(2b + 1). \tag{S31}$$

In the general case of $n_i$ electrons per pulse, it can be shown (through a similar demonstration as in section S2d)

$$A_{n_i} = n_i b m^2 \eta^2 (b n_i + 1). \tag{S32}$$

Finally, the number of electrons per pulse is not fixed but follows a Poisson distribution with expected value $n_e$. Moreover, we need to multiply this by the total number of pulses exciting the sample during a measurement ($r$). Hence, the average number of combinations of correlations leading to bunching becomes

$$A'_b = rbm^2\eta^2 \sum_{n_i=0}^{\infty} n_i(bn_i + 1)\text{Poiss}(n_i; n_e) = rn_e b m^2 \eta^2 (n_e b + b + 1). \tag{S33}$$

The area below each peak $i$, containing the number of combinations of correlations between photons from different pulses (consecutive pulses, every second pulse, etc) was already calculated in Eq. (S27). Hence,

$$A'_i = A^p_{\text{uncorr},i} = \frac{2 A^p_{\text{uncorr,r}}}{2(r-1)} = rn_e^2 b^2 m^2 \eta^2. \tag{S34}$$

Inserting Eqs. (S33) and (S34) into Eq. (S30) yields

$$g^{(2)}_{\text{ultrashort,v2}}(0) = 1 + \frac{b+1}{n_e b}, \tag{S35}$$

which is the same as $g^{(2)}_{\text{ultrashort,}}(0)$ given in Eq. (12) of the main text, which was obtained by setting $\alpha_{conv} = \alpha_\text{b}$ in Eq. (S28).

## S4. Full description of $g^{(2)}(\tau)$

In the previous sections we have derived the value of $g^{(2)}(0)$, but we have not discussed yet the full shape of the autocorrelation function as a function of delay (i.e., $g^{(2)}(\tau)$). In the continuous case, the shape of $g^{(2)}(\tau)$ only depends on the bunching peak, while in pulsed experiments $g^{(2)}(\tau)$ depends only on the temporal shape of the electron pulses, as will be seen below.

### S4a. Shape of the bunching peak

Given a certain function $y(t)$, the result of its autocorrelation is[4]

$$h(\tau) = \int_{-\infty}^{\infty} y(t)y(t+\tau)\,dt = y(-\tau) * y(\tau). \tag{S36}$$

In the case of the bunching peak in a $g^{(2)}(\tau)$ measurement, $y(t) = y_{\text{emitter}}(t)$ and $h(\tau) = y_{\text{bunching}}(\tau)$, as given in Eq. (7) in the main text.

### S4b. Shape of uncorrelated peaks in a pulsed electron beam

As we have already discussed, a $g^{(2)}(\tau)$ measurement in pulse shows peaks centered 0-delay and delays $\tau_i$ ($i = \pm 1, \pm 2 \ldots$) corresponding to the time between pulses. The peak at $\tau = 0$ has contributions from bunching, which result in a shape determined by the emitter as shown in Eq. (S35) and Eq. (9) in the main text, and from uncorrelated photons, i.e., coming from different electrons. The peaks at $\tau_i$ ($i \neq 0$) contain also uncorrelated photons, i.e., coming from different pulses. In all cases in which there are correlations between photons from different electrons, the shape of the electron pulse also plays a role, together with the emitter decay. The probability of emitting a photon coming from a pulsed electron beam is given by the convolution between the electron pulse shape ($p(t)$) and emitter decay ($y_{\text{emitter}}$), i.e.,

$$y(t) = p(t) * y_{\text{emitter}}(t). \tag{S37}$$

And from Eq. (S35), we know that the correlation between two photons with temporal spread $y(t)$ is

$$h^p_{\text{uncorr}}(\tau) = [p(-\tau) * y_{\text{emitter}}(-\tau)] * [p(\tau) * y_{\text{emitter}}(\tau)] = [p(-\tau) * p(\tau)] * h_\text{b}(\tau), \tag{S38}$$

where in the last step we have used the definition of $h_\text{b}(\tau)$ from Eq. (7) in the main text.

### S4c. Comparison to experiments: stretched exponential decay

In order to test the validity of Eq. (7) (main text) (same as Eq. (S35)), describing the shape of the bunching peak, we performed time-correlated single-photon counting measurements (TCSPC) on the sample. We subsequently acquired a $g^{(2)}(\tau)$ measurement with exactly the same conditions. The TCSPC measurements were performed using a pulsed electron beam obtained by photoemission, with the same conditions as in the $g^{(2)}(\tau)$ photoemission experiments (here, $n_e = 347$ electrons/pulse), and the data is collected in the same way as explained in refs.[6,7]. In these measurements, a histogram of the arrival time of photons following the electron pulse is built, and thus they directly show the emission decay. Figure S3a shows the resulting decay trace. We observe that the trace can be best fitted using a stretched exponential

$$y_{emitter}(t) = y_0 e^{-\left(\frac{t}{\tau_{emitter}}\right)^{\beta_{emitter}}}. \tag{S39}$$

In this case we find that $\tau_{emitter} = 11$ ns and $\beta_{emitter} = 0.73$. The histogram obtained in the corresponding $g^{(2)}(\tau)$ measurement is shown in Figure S3b. We observe that the bunching peak cannot be properly described with a stretched exponential using $\tau_{emitter}$ and $\beta_{emitter}$ as parameters (green curve). Instead, the result of solving numerically Eq. (S35) with the emitter parameters exhibits a very good agreement with the data (red curve).

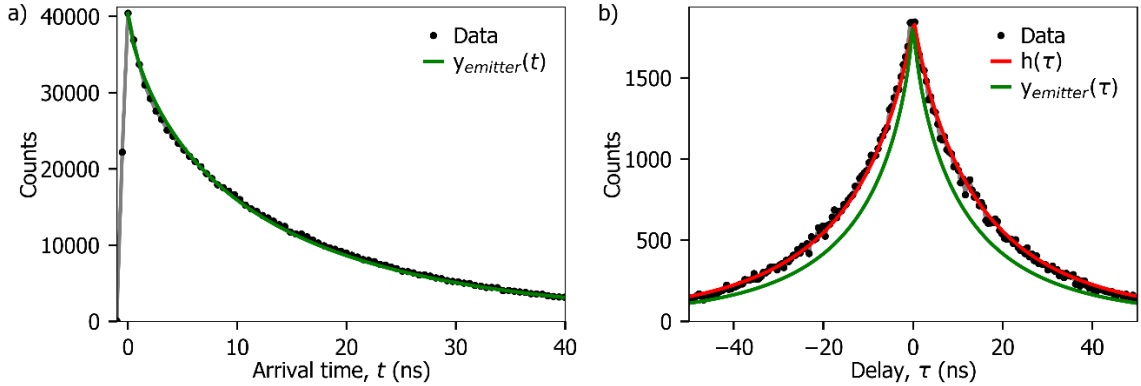

**Figure S3:** (a) TCSPC measurement on quantum wells, performed using a pulsed electron beam generated by photoemission (black) together with the corresponding fit with a stretched exponential (green). The best fit is obtained when $\tau_{emitter} = 11$ ns and $\beta_{emitter} = 0.73$. (b) $g^{(2)}(\tau)$ measurement performed on the same area on the sample and identical conditions as in (a) (black), together with the result from Eq. (S35) using the emitter parameters from (a) (red) and a stretched exponential using the emitter parameters. The model (Eq. (S35)) shows a very good agreement with the data.

The discrepancy between the shape of the $g^{(2)}(\tau)$ curve and the actual emitter decay when the latter follows a stretched exponential could explain the different lifetimes obtained in ref. [7] when comparing $g^{(2)}(\tau)$ and decay trace measurements.

*Single and double exponential decays*

In most systems, the decay mechanism can be approximated with a single or double exponential decay. Solving Eq. (S35) in those cases yields: $y_{bunching}(\tau) \propto e^{-\tau/\tau_1}$ and $y_{bunching}(\tau) \propto e^{-\frac{\tau}{\tau_1}} + e^{-\tau/\tau_2}$, for single and double exponential decays, respectively. Therefore, in both cases the decay of the $g^{(2)}(\tau)$ function directly gives the decay of the emitter.

### S4d. Comparison to experiments: square electron pulse

The experiments using the beam blanker are performed using square electron pulses, with pulse width $\Delta p$ determined by the blanking conditions (repetition rate and duty cycle) (see Section S6a). Hence, the pulse shape is given by

$$\tag{S40}$$

$$p(t) = \begin{cases} 1, & 0 \leq x \leq \Delta p \\ 0, & \text{otherwise,} \end{cases}$$

and the emitter decay $y_{\text{emitter}}(t)$ follows the expression in Eq. (S38). The shape of the peaks at $\tau_i (i \neq 0)$ then become (Eq. (S37))

$$h_{\text{uncorr}}^p(\tau) = T(\tau) * [y_{\text{emitter}}(-\tau) * y_{\text{emitter}}(\tau)] = T(\tau) * h_b(\tau), \tag{S41}$$

where $T(\tau)$ is a triangular function with base $\Delta p$, resulting from the convolution of $p(t)$ with $p(-t)$.

## S5. Correction at long delays

In a $g^{(2)}(\tau)$ measurement, when the delay is longer than the typical correlation time (in our case, the emitter lifetime), we expect all events to be uncorrelated, thus exhibiting a constant amplitude. In the case of a continuous electron beam, this means that the $g^{(2)}(\tau)$ curve is constant for $\tau \gg 0$, while in the pulsed case, we still observe peaks at the delays corresponding to the time between pulses, all of them with the same amplitude. Nevertheless, this is not typically what we observe in experiments. Figure S4 shows the raw data of two $g^{(2)}(\tau)$ measurements, in continuous (a) and pulsed (b) mode. In both cases we observe that the number of counts decreases with increasing $\tau$, contrary to what we would expect from the theory. This is due to an experimental artifact in the Hanbury-Brown and Twiss experiment. In the experiment, the emitted light is split into two beams with a 50:50 beam splitter. Each beam is directed towards one detector, connected to the time correlator. When one of the detectors receives a photon, the time correlator starts counting until a photon is received on the second detector. Therefore, having a count at a certain delay $\tau$ means that the second detector does not receive any photon during the time $\tau$. This becomes very unlikely with increasing $\tau$, thus producing the effect observed in the figure.

One way to avoid this artifact is by having a very low count rate on each detector, such that the probability of having two (uncorrelated) photons emitted within a time smaller than $\tau$ becomes very low. Nevertheless, this can result in very long acquisition times (in the order of hours) or low signal-to-noise ratios. In our case, we decided to keep the number of counts relatively high (typically $10^4$ counts/s) and correct for this artifact during the data analysis. We observe that the evolution of the signal over $\tau$ due to this artifact follows an exponential decay, with average decay $\tau_{long}$. The fits obtained when applying this decay are shown in Figure S4, for which we obtained $\tau_{long} = 31$ and 769 μs, respectively. This procedure is valid as long as $\tau_{long}$ is much larger than the bunching decay and pulse width, in the case of a pulsed electron beam. Otherwise, artifacts due to this effect would also affect the value of $g^{(2)}(0)$.

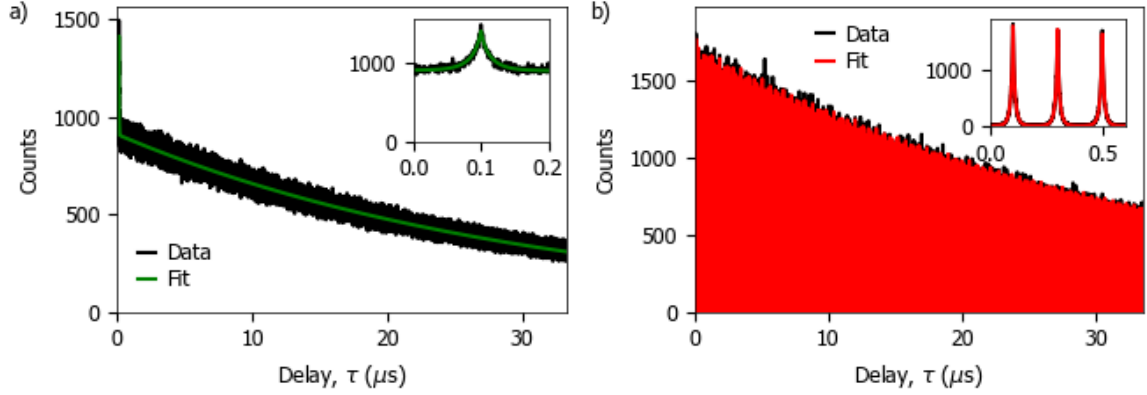

**Figure S4:** Raw data from a $g^{(2)}(\tau)$ experiment with a (a) continuous and (b) pulsed (by photoemission) electron beam. The data show that the number of counts decrease for long delays due to an experimental artifact in the HBT experiment. The green and red curves show the fits using an exponential decay to account for this artifact. The obtained average decays are $\tau_{long} = 31$ and $769\ \mu s$ for (a) and (b), respectively. The insets show a zoom in for small delays, in which this artifact is not visible.

## S6. Experimental details

All measurements are performed when focusing the electron beam on a single spot on the sample. The electron current is measured by collecting the beam current through a Faraday cup and reading it with a picoammeter.

### S6a. Beam blanker

The experiments using a beam blanker are performed using the same microscope as in ref. [7]. In our case, a 400 μm aperture is placed right below the pole piece. The distance between the blanking plates is kept to 2 mm for all experiments. In contrast to previous work, here we apply a square signal on one of the blanking plates, with peak-to-peak amplitude of 5V and offset 2.5V. The other plate is grounded. This results in a square electron pulse, with pulse width determined by the duty cycle $D$ and repetition rate $F$, i.e., $\Delta p = (1-D)/F$. In order to confirm the shape of the electron pulse, we performed decay trace measurements on the QWs while blanking the beam. Figure S5 shows two examples of traces, both obtained using $D = 0.6$ and repetition rate $F = 0.5$ and 6 MHz, respectively. We fitted the data using the following equation

$$f(x) = \begin{cases} B, & t < t_0 \\ A\left(1 - e^{-\left(\frac{(t-t_0)}{\tau}\right)^\beta}\right) + B, & t_0 \leq t \leq t_1 \\ Ae^{-\left(\frac{(t-t_1)}{\tau}\right)^\beta} + B, & t > t_1, \end{cases} \quad (S42)$$

where $\tau = 8.6$ ns and $\beta = 0.63$ are the parameters describing the QW radiative decay, $A$ is the amplitude of the signal and $B$ is the background signal. The pulse width can be obtained from $\Delta p = t_1 - t_0$. In the experiments we obtain pulse widths of 796 and 62 ns, for Figure S5(a) and (b) respectively, which are very close to the theoretical values at these conditions (800 and 66 ns, respectively). These experiments were performed using an electron energy of 10 keV, but we do not expect significant deviations when changing the electron energy to 8 keV.

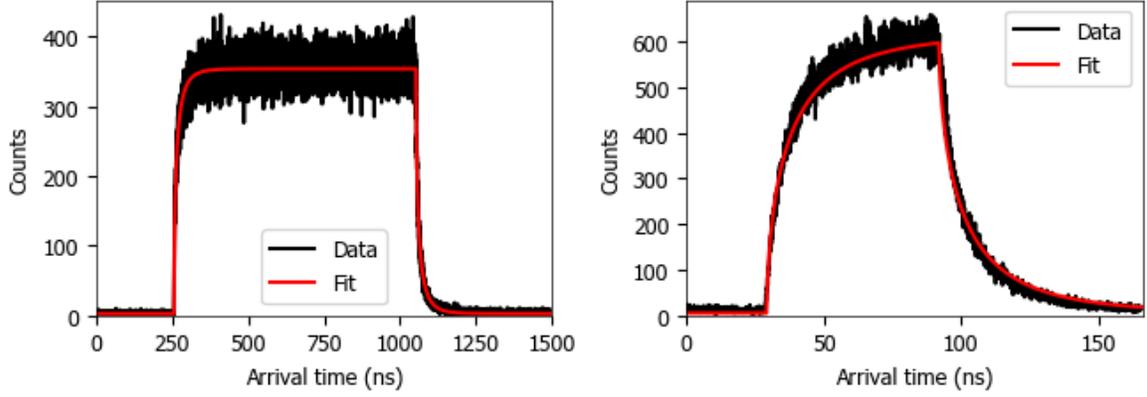

**Figure S5:** Decay traces on the QWs obtained using an electron beam blanker, with repetition rate (a) 0.5 and (b) 6 MHz. The fits are obtained using Eq. (S41), showing that the electron pulses can be described as a square pulse with pulse width of 796 and 62 ns, for (a) and (b) respectively.

Even though the experimental data shows an almost perfect square electron pulses, small deviations from this can arise when changing parameters, especially when increasing the duty cycle and repetition rate. In order to account for this, we measured the electron current in continuous mode $I_c$ (i.e., in blanking conditions but without any signal driving the blanking plates) and in pulsed $I_p$ (square signal driving one of the plates). The relation between both magnitudes is given by $I_p = I_c(1 - D)$. Figure S6a shows the value of electron current in pulsed $I_p$ measured at different repetition rates. These measurements were performed at 8 keV and $D = 0.95$, with the same blanking conditions as for the $g^{(2)}(\tau)$ measurements using the blanker in the main text. The figure also shows the expected value of $I_p$ (red curve), given a continuous current of $I_c = 213.9$ pA. We observe that the measured values are slightly lower than the expected ones, and the discrepancy increases with increasing repetition rate. These measured values of $I_p$ were used to calculate the number of electrons per pulse in Fig. 3 of the main text. The pulse duration of the electron beam can also be extracted from these measurements, given that $\Delta p = I_p / I_c F$. Figure S6b shows the value of pulse width obtained using the experimental values of $I_p$ (black dots) compared to the theoretical values, given by $\Delta p = (1 - D)F$ (red curve).

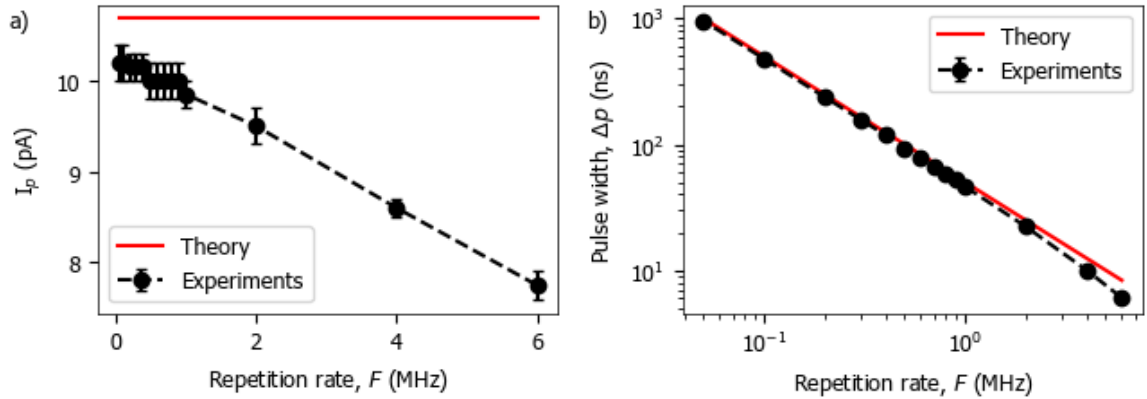

Figure S6: (a) Electron beam current measured in pulsed conditions using a beam blanker as a function of repetition rate. The red curve represents the theoretical current that we should obtain given a continuous current of 213.9 pA, and duty cycle of 0.95. (b) Pulse duration extracted from the experimental values of electron current in pulsed in (a), together with the theoretical value of the pulse width (red curve).

### S6b. Laser-driven electron source (photoemission)

Ultrashort (ps) pulses are obtained by focusing the 4th harmonic (257 nm) of an Yb-doped femtosecond laser ($\lambda = 1035$ nm, 250 fs pulses) onto the electron cathode. The experiments are performed using a Quanta 250 FEG SEM. In order to suppress continuous emission, the filament current is reduced from 2.35 down to 1.7 nA. The extractor voltage is also lowered from the typical 4550V value down to 650V. These settings allow us to achieve a high number of electrons per pulse, at the expense of lower spatial resolution, as explained in ref [7].

## S7. Cathodoluminescence with 8 keV electrons

Figure S7 shows the CL spectrum obtained when exciting the sample with a continuous 8 keV electron beam, corresponding to the energy used in the experiments using the beam blanker. Most of the emission comes from the QW emission (410-490 nm). The inset shows a schematic of the structure of the sample together with Casino simulations for an 8 keV electron. Each dot in the plot corresponds to an inelastic collision of the primary electron beam with the sample, while the color indicates the energy of the primary electron beam. We observe that barely any electron reaches the QWs, thus explaining the low excitation efficiency obtained at 8 keV ($\gamma = 0.05$).

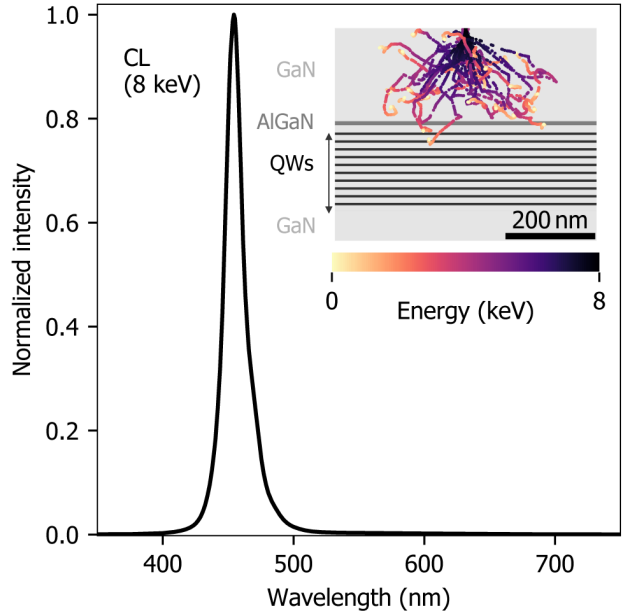

**Figure S7:** CL spectrum obtained after excitation with an 8 keV continuous electron beam (213.5 pA). Inset: schematic of the InGaN/GaN quantum well stack overlaid with Casino simulations

## S8. Dependence of QW emission decay on area

In the main text we show $g^{(2)}(\tau)$ measurements performed using the different electron beam configurations (continuous, pulsed with blanker and pulsed with photoemission), in each case exhibiting different decay lifetimes ($\tau_{emitter}$ and $\beta_{emitter}$). Here we prove that the main reason for this discrepancy is the inhomogeneity in the sample. Figure S8 shows $g^{(2)}(\tau)$ measurements performed using a continuous electron beam on different spots on the sample. The curves were obtained at 10 keV with beam currents of 10.6, 14.1 and 34 pA (green, blue and yellow curves, respectively). Each experimental curve (data points) is accompanied by the corresponding fit (solid lines), obtained by solving numerically Eq. (S35) when $y(t)$ is a stretched exponential. We observe that $\tau_{emitter}$ strongly depends on the position of the sample, ranging from 3.7 to 7.3 in these three examples. Instead, $\beta_{emitter}$ remains in the 0.61-0.64 range.

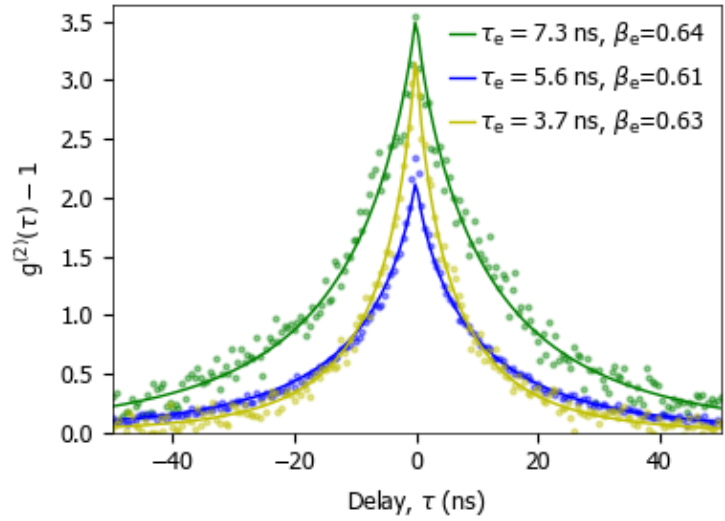

**Figure S8:** $g^{(2)}(\tau)$ measurements obtained with a 10 keV continuous electron beam at three different spots on the sample. The solid lines are the fits from Eq. (S35) when $y(t)$ is a stretched exponential, with fit parameters $\tau_e \equiv \tau_{\text{emitter}}$ and $\beta_e \equiv \beta_{\text{emitter}}$.